\documentclass{optica-article}

\journal{opticajournal} 

\articletype{Research Article}

\usepackage{lineno}

\usepackage{multicol} 
\usepackage{multirow} 

\newcommand{\N}{\mathbb{N}}

\newcommand\norm[1]{\left\lVert#1\right\rVert}
\newcommand\abs[1]{\lvert#1\rvert}

\begin{document}

\title{Effective stripe artifact removal by a variational method: application to light-sheet microscopy, FIB-SEM and remote sensing images}

\author{Niklas Rottmayer\authormark{1,*}, Claudia Redenbach\authormark{1} and Florian O. Fahrbach\authormark{2}}

\address{\authormark{1} Mathematics Department, RPTU Kaiserslautern-Landau, Gottlieb-Daimler-Straße, 67663 Kaiserslautern, Germany\\
\authormark{2} Leica Microsystems CMS GmbH, Am Friedensplatz 3, 68165 Mannheim, Germany}


\begin{abstract*} 
Light-sheet fluorescence microscopy (LSFM) is used to capture volume images of biological specimens. It offers high contrast deep inside densely fluorescence labelled samples, fast acquisition speed and minimal harmful effects on the sample. However, LSFM images often show strong stripe artifacts originating from light-matter interactions. We propose a robust variational method suitable for removing stripes which outperforms existing methods and offers flexibility through two adjustable parameters. This tool is widely applicable to improve visual quality as well as facilitate downstream processing and analysis of images acquired on systems that do not provide hardware-based destriping methods. An evaluation of methods is performed on LSFM, focused ion beam scanning electron microscopy (FIB-SEM) and remote sensing data, supplemented by synthetic LSFM images. The latter is obtained by simulating the imaging process on virtual samples. 
\end{abstract*}

\section{Introduction}
\begin{figure}[htbp]
    \centering
    \includegraphics[width=\textwidth]{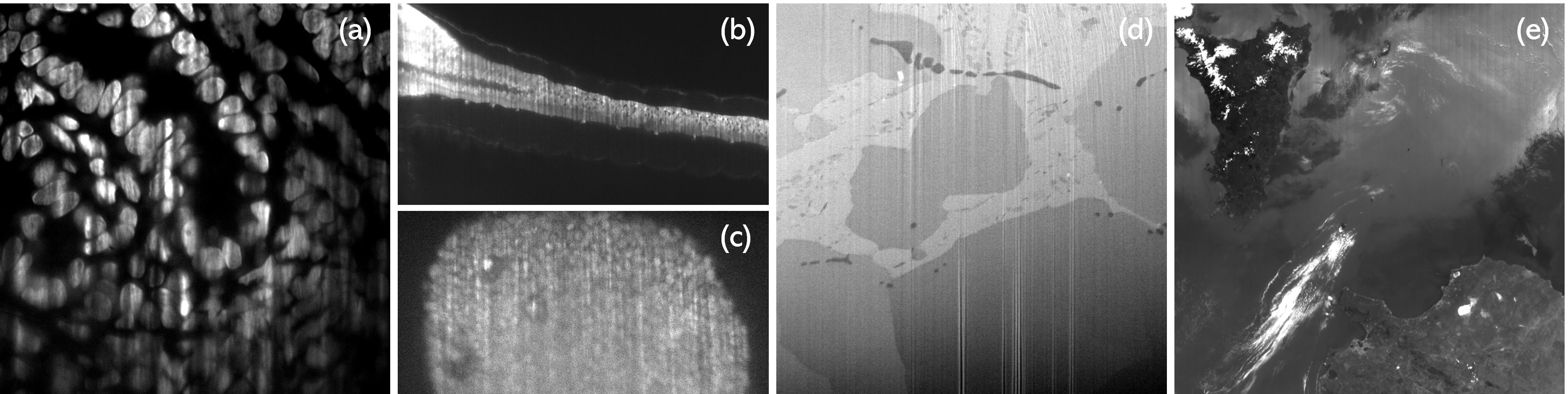}
    \caption{Images corrupted by stripe artifacts from LSFM (a-c), FIB-SEM (d) and remote sensing (e). Cultivated cell organoid of mouse intestine cells (a), tail of Zebrafish larva (b), cluster of HeLa cells (c), image slice of a tin bronze cast alloy (d) and Terrra MODIS data (e).}\label{fig:StripeExamples}
\end{figure}

The observation by microscopy plays a fundamental role in understanding multicellular life and biological processes. Since the beginning of this century, light-sheet (fluorescence) microscopy (LSFM) has gained popularity in 3D-imaging of biological specimens. It provides high resolution and fast acquisition speeds \cite{Huisken2004} while minimizing harmful exposure to light \cite{Jemielita2013}. However, the acquired images are often accompanied by strong stripe artifacts caused by light absorption and scattering, see Fig. \ref{fig:StripeExamples} (a-c). The corruptions can occur in large parts of the imaged volume and obscure underlying structures, thus impairing their appearance and complicating further data analysis. There is a large quantity of articles concerned with stripe artifacts in LSFM \cite{Fahrbach2010,Huisken2007,Keller2010,Liang2016,Rohrbach2009,Salili2018,Scherf2015}. In a recent paper, Ricci and colleagues \cite{Ricci2022} present a detailed summary of previous research on removing stripes from LSFM images. They discuss hardware solutions to prevent stripe formation, algorithms to remove stripes through post-processing and hybrid methods which combine ideas from both worlds.   
		
At first glance, the prevention of artifacts is clearly the preferential choice. However, this requires modifications of the imaging setup which are costly and often require entirely new microscopes. They may also limit the acquisition speed, the light-efficiency or be incompatible with advantageous image acquisition modes such as line-confocal acquisition \cite{Fahrbach2012,Fahrbach2013a,Fahrbach2013b}. In comparison, algorithmic post-processing is cheap and requires only computational power. Furthermore, corrupted images may already exist and specimens cannot be re-imaged, e.g., due to degradation and aging. For these reasons it is desirable to have powerful post-processing tools for stripe removal available.  
    
Stripe artifacts are not unique to LSFM. They are also common in imaging techniques such as atomic force microscopy (AFM) \cite{Chen2011}, focused ion beam scanning electron microscopy (FIB-SEM) \cite{Fitschen2017} or remote sensing \cite{Liu2018,Chang2016}, see Fig. \ref{fig:StripeExamples} (d-e) for examples. Although having different causes, stripes are very similar in shape showing pronounced elongated structures in a common direction with mostly similar thickness of only a few pixels. Their shape deviates strongly from underlying structures and presents a feature that can be used for detection and removal. In contrast, the appearance of stripes in LSFM is much more diverse. While stripes are still aligned along a common direction, their length, thickness and intensity can vary drastically depending on the imaged structures. For example, a single stripe can be split into smaller separated segments of lengths equal to that of underlying structures, see Fig. \ref{fig:StripeExamples} (a). Additionally, stripes display a much larger range of thickness and depth such that individual stripes can appear in multiple slices of 3D volume images. This drastically complicates the correct detection and removal of such artifacts, as stripe removal methods heavily rely on the distinctive shape differences to image structures. 

In this paper, we discuss the two major categories of stripe removal methods in research, Fourier filtering and variational methods. We explain the functionality of Fourier filtering methods on the example of the multi-directional stripe remover (MDSR) \cite{Liang2016} and present a modification to increase its performance. For the variational methods we propose a model which is based on the work of Liu and colleagues \cite{Liu2018} and performs exceptionally well on all data tested. Furthermore, we suggest possible adaptations to make it applicable to a broad range of stripe removal settings. We evaluate and compare the performance of the methods using real and synthetic image data. The latter is obtained through physically correct simulation of light transport through randomly generated samples and provides a stripe-free reference image. The simulation is achieved using the Python package biobeam \cite{Weigert2018} and, to our knowledge, has not previously been used to validate stripe removal. The availability of a ground truth, which is unavailable in experimentally acquired images, allows for an objective validation of stripe removal methods as quality metrics such as the peak signal-to-noise ratio (PSNR) or the multi-scale structural similarity index measure (MS-SSIM) can be applied. For the evaluation of real and synthetic data we rely on visual inspection and the curtaining metric proposed by Rold\'an \cite{Roldan2024} which measures the amount of corruptions by stripes and does not require reference images.

\section{Methods}\label{sc:Methods}
The majority of methods proposed for stripe removal fall under the two categories \textit{Fourier filtering} and \textit{variational methods}. We will concentrate on these categories, as they generalize well to several imaging methods and variations in image structures and stripe appearance. Additional approaches are average filtering \cite{Ashrafuzzaman2010,Boin2006}, histogram matching \cite{Cao2015,Rakwatin2007}, spline interpolation \cite{Tsai2005} and recently neural networks \cite{Liu2022,Wei2022,Wang2023,Pietsch2021}. However, these are usually tailored for a specific appearance of images and stripes such that they are harder to transfer to other scenarios.

\subsection{Fourier Filtering}\label{ssc:FourierFiltering}
\begin{figure}[htbp]
    \centering
    \includegraphics[width=1\textwidth]{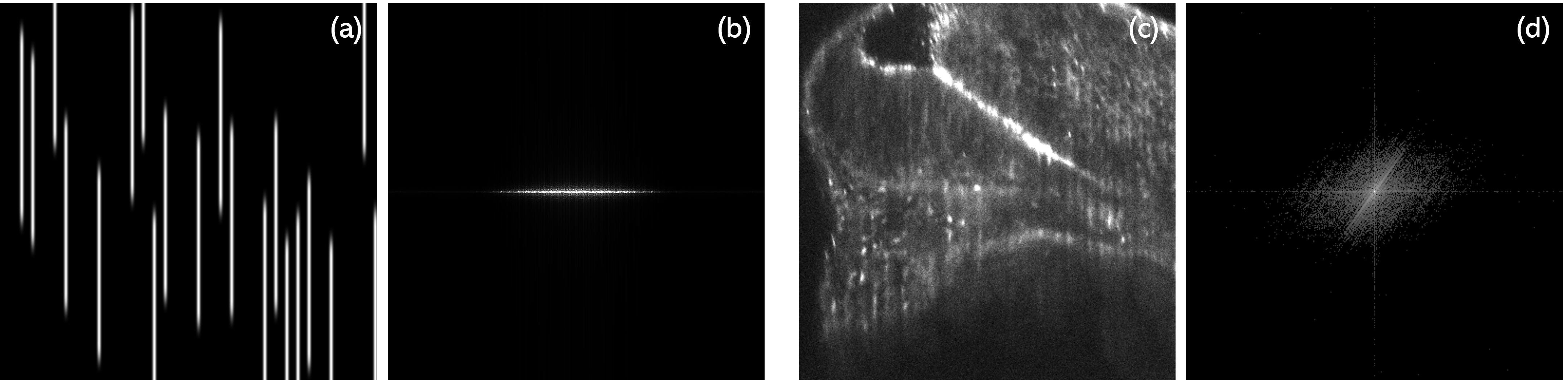}
    \caption{Image and spectrum of thin stripes (a-b) and an LSFM image (c-d).}\label{fig:FourierStripes}
\end{figure}

The Fourier transform yields a mathematical decomposition of the image content into constituent frequencies. 
In image processing, the breakdown into a frequency spectrum is used to detect and modify periodic structures of certain frequencies or entire bands of frequencies. 
We assume that stripe artifacts point into the same direction, have small widths and appear repeatedly. In the Fourier domain, stripe information is encoded in frequency coefficients around a small band orthogonal to the stripe direction. In contrast, natural structures 
live on larger scales than the widths of stripes. Thus, image structure information is dominated by low frequencies which are concentrated in central coefficients. This general behavior is visualized in Fig. \ref{fig:FourierStripes}.

\begin{figure}[htbp]
    \centering
    \includegraphics[width=\textwidth]{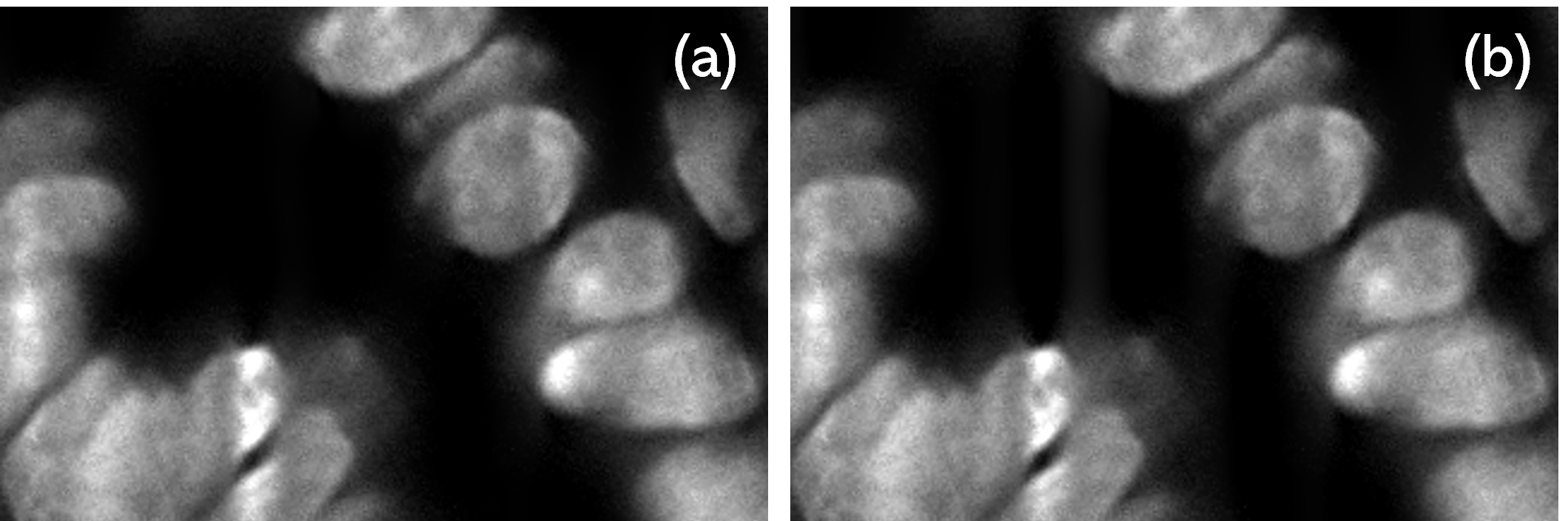}
    \caption{Visual comparison of restricting Fourier filtering to subimages of directions $\theta_i\in[\frac{\pi}{4},\frac{3\pi}{4}]$ (a) to unrestricted filtering \cite{Liang2016} (b). Restricting filtering reduces artifacts introduced in dark areas while maintaining a similar level of stripe removal.}\label{fig:MDSRComparison}
\end{figure}

By appropriate damping or removal of  coefficients attributed to stripe artifacts, corruptions can be significantly reduced. 
This can be achieved, e.g., by using masked filters \cite{Wilken2012,Liu2016a,Schwartz2019} or a decision-based algorithm \cite{Chen2011}. A prior decomposition by structure and scale through wavelets \cite{Muench2009} or the non-subsampled contourlet transform (NSCT) \cite{Cunha2006} reduces the effect of stripe removal on the image structure. 
We will use the multi-directional stripe remover (MDSR) \cite{Liang2016} based on the NSCT. However, we use a slightly modified filtering step to improve quality. The MDSR depicts an improvement over wavelet Fourier filtering proposed by Münch et al. \cite{Muench2009} which is commonly referenced and one of the earliest successful adoptions of Fourier filtering for removing stripes. The workflow of MDSR is visualized in Fig. \ref{fig:MDSR} and consists of the following steps:

\begin{enumerate}
\item \textbf{Image Decomposition:}
Initially, stripe and image information are partially separated by application of the NSCT. Using a pyramidal filter bank \cite{Cunha2006}, the image is decomposed into subimages depending on direction and scale until a selected depth $n_{\mathrm{dec}}\in\N$ is reached. The number of decomposition directions is denoted by $n_{\mathrm{dir}}\in\N$.  

\item \textbf{Fourier Filtering:}
    Subsequently, coefficients are filtered on subimages in the Fourier domain. 
    Contrary to the original MDSR \cite{Liang2016}, we apply damping only to subimages of directions with $\vert\theta_i-\theta_0\vert\leq\pi/4$. This reduces artifacts as shown in Fig. \ref{fig:MDSRComparison}. For vertical stripes, i.e., $\theta_0=\pi/2$, damping is performed via element-wise multiplication with
    \begin{align*}  f_i(x,y) =\begin{cases}
        1-\exp\left(-\frac{(y-\frac{n_y}{2})^2}{2\sigma_i^2}\right) & \text{if } \vert\theta_i-\theta_0\vert\leq\pi/4, \\
        1 & \text{otherwise},
    \end{cases}, \quad 
    \sigma_i=\sigma\cdot\exp\left(-\frac{(\theta_0-\theta_i)^2}{2\sigma_a^2}\right),
    \end{align*}
    with Gaussian standard deviation $\sigma$ and vertical image size $n_y$. The damping parameters $\sigma_i, \, i=1, \ldots, n_{\mathrm{dir}},$ depend on the deviation of  $\theta_i$ from the stripe direction $\theta_0$. The parameter $\sigma_a>0$ describes how quickly damping is reduced when moving away from the stripe direction.

\item \textbf{Backtransformation:}
    The modified Fourier coefficients are transformed back into individual subimages. NSCT reconstruction then yields the destriping result.     

    \begin{figure}[htbp]
        \centering
            \includegraphics[width=0.9\textwidth]{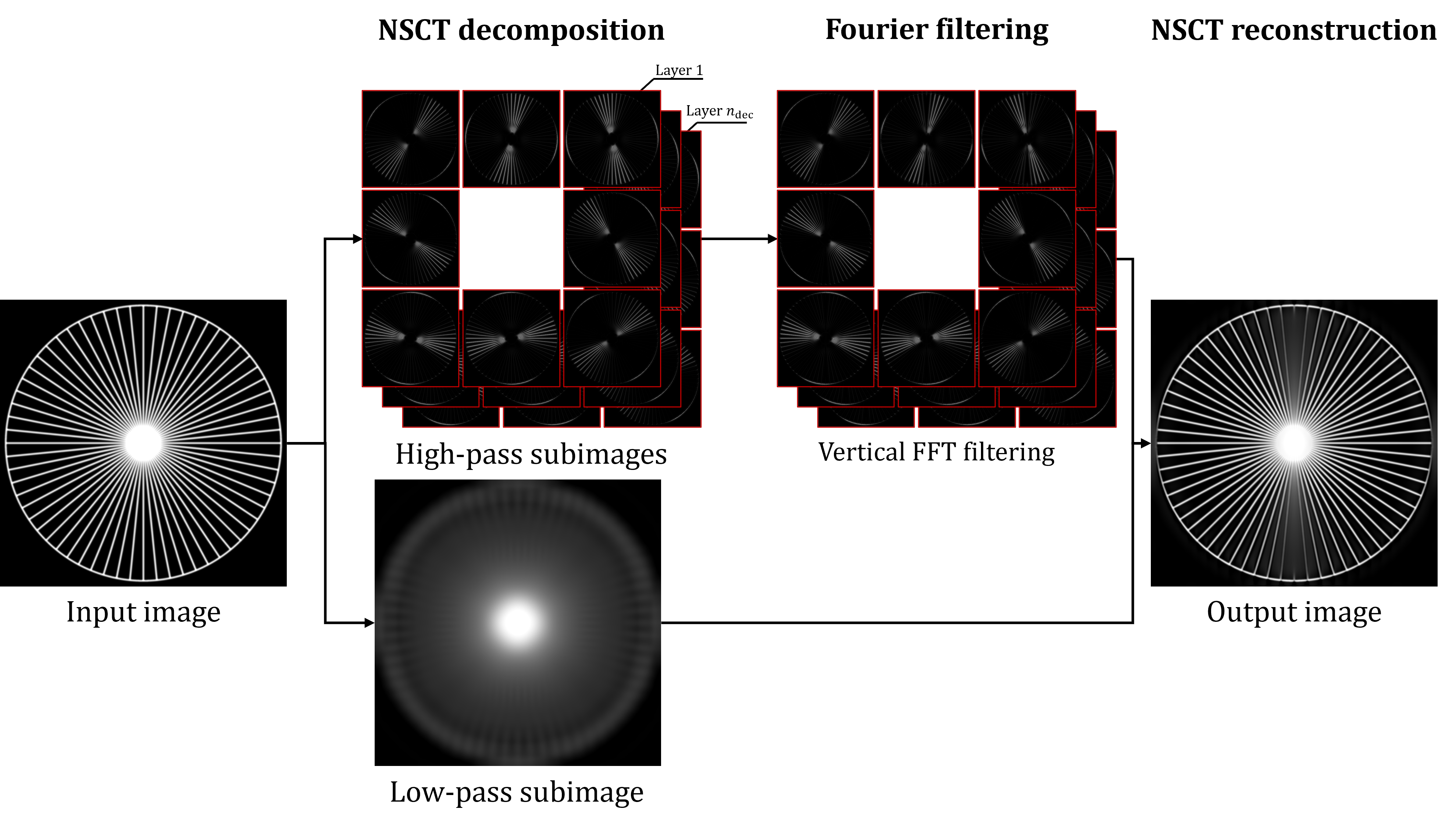}
        \caption{MDSR workflow for removing vertical stripes.}\label{fig:MDSR}
    \end{figure}
\end{enumerate}

The MDSR algorithm depends on multiple parameters: the number of directions $n_{\mathrm{dir}}$, the decomposition depth $n_{\mathrm{dec}}$, a Gaussian damping parameter $\sigma$, a directional fall-off parameter $\sigma_a$ and the selected filter banks. However, most of them do not require fine-tuning once chosen adequately. For example, $n_{\mathrm{dir}}=8$ yields a sufficient directional decomposition and $n_{\mathrm{dec}}$ must only be large enough such that stripes are captured by the NSCT. Any further increase in both parameters has a negligible effect on the outcome while significantly increasing the computational complexity. $\sigma$ is the main parameter to be adjusted and strongly influences the outcome as it depicts the strength of damping. When stripe artifacts are not ideally vertical but slightly oblique, adjusting $\sigma_a$ may also improve performance.  

Fourier filtering approaches such as MDSR rely on separation of stripe information and image content in the Fourier domain. 
However, stripes in LSFM images cover a wide range of lengths, widths and intensities such that stripe information is less concentrated around the horizontal coefficient band and more mixed up with image information. This complicates stripe removal. 

\subsection{Variational Methods}\label{ssc:VariationalMethods}
Variational methods achieve destriping by minimizing  a problem-specific function consisting of a weighted sum of convex functions capturing multiple aspects of the stripe removal problem.
Under mild assertions, convex functions have unique minimizers which can be approximated using well-studied optimization algorithms, see \cite{Burger2016}. 
The individual functions penalize unwanted features of the clean image $u$ and stripes $s$ through positive contribution to the function. A corresponding minimizer should then reasonably fulfill the desired properties, i.e.,
optimizing the objective function yields a solution to the problem. This approach is well known in image processing and commonly used for denoising, segmentation and active contours \cite{Vese2016}. 
We adjusted a functional originally proposed by Fitschen et al. \cite{Fitschen2017} for curtaining noise in FIB-SEM images to the LSFM setting. The result shows resemblance to the proposition by Liu et al. \cite{Liu2018} but is also applicable to 3D images and includes an additional regularization term which improves performance and consistency.

Besides our proposition, there exist a variety of alternative objective functions. An early and more general proposition comes from Fehrenbach and colleagues \cite{Fehrenbach2012} designed to remove repeating artifacts. The model relies on the representation of artifacts 
in terms of elementary noise patterns which makes it applicable to  any kind of noise created by repeating specific patterns. In a follow-up paper by Escande et al. \cite{Escande2017}, multiplicative noise of the same form is considered. This represents stripe formation in LSFM more closely. However, the authors have noticed a significant change in contrast as side effect of their method which can strongly alter the image appearance. 
Other propositions include \cite{Chang2016,Chen2022,Yan2023,Zhou2023} utilizing sparsity and low-rank assumptions on $s$ 
and \cite{Gourtani2019} with a combination of Fourier filtering and variational methods. 


\subsubsection{Construction}
Consider the corrupted image $u_0\in[0,1]^N$, the clean image $u\in[0,1]^N$ and the stripe image $s\in[-1,1]^N$ where $N=n_x\times n_y\times n_z$. We assume that $u_0 = u+s$. Let $\nabla_x,\nabla_y$ and $\nabla_z$ denote the directional difference operators for the three coordinate directions and assume that stripes point in $y$-direction.   

\begin{figure}[htbp]
    \centering
    \includegraphics[width=0.8\textwidth]{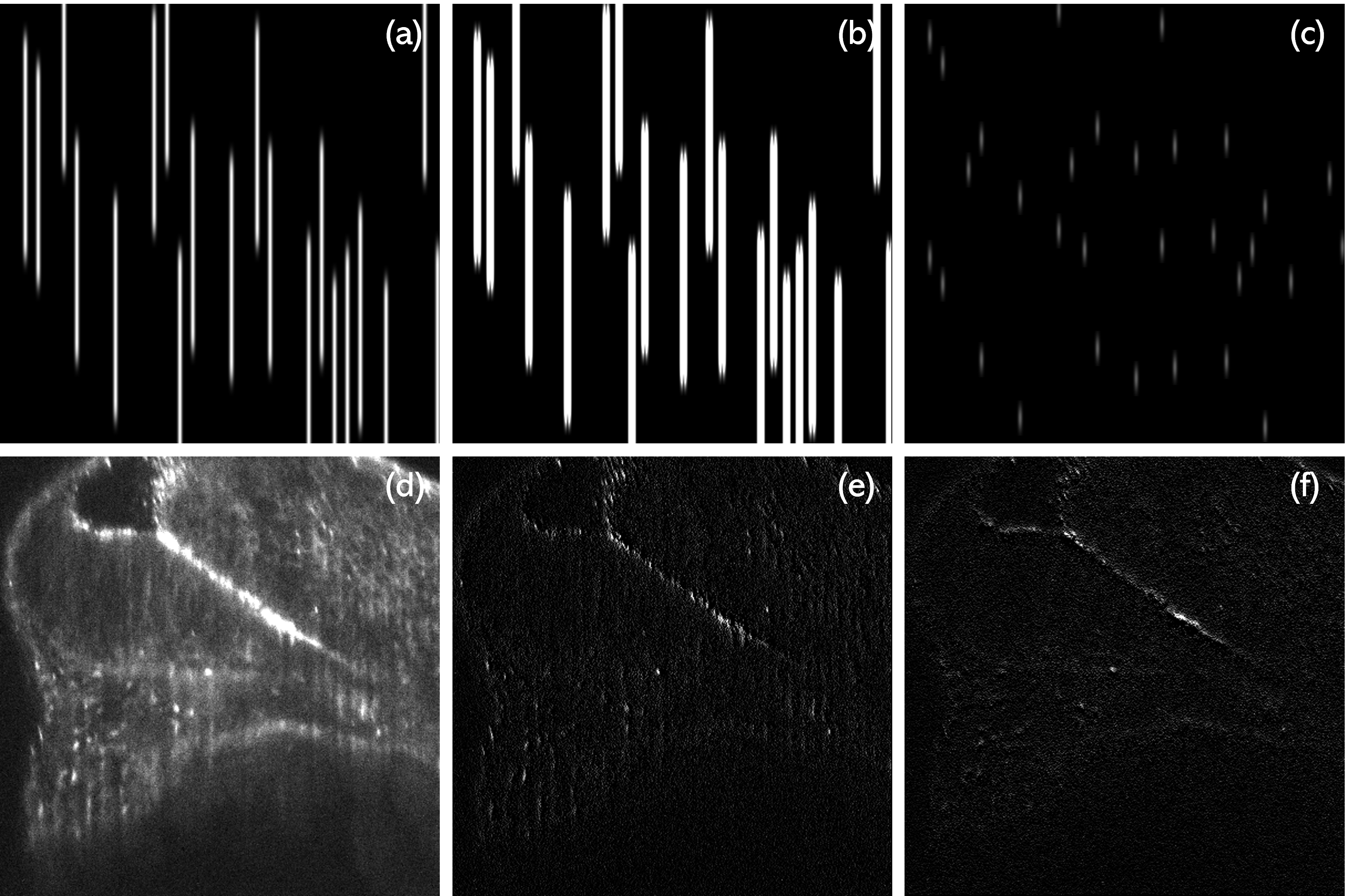}
    \caption{Images and directional differences $\nabla_x,\nabla_y$ of thin stripes (a-c) and LSFM image (d-f).}\label{fig:VarMethodsMotivation}
\end{figure}

The objective function is grouped into a data term $D(u)$ and a noise term $N(s)$ such that the general optimization problem reads
\begin{equation}
    \underset{u+s=u_0}{\mathrm{argmin}}\ D(u) + N(s).
\end{equation}  
The \textit{total variation}
\begin{equation} 
    \norm{\nabla u}_{2,1} = \sum_{i,j,k}\sqrt{(\nabla_{x} u)_{i,j,k}^2 + (\nabla_{y} u)_{i,j,k}^2 + (\nabla_{z} u)_{i,j,k}^2}
    \label{eq:TotalVariation}
\end{equation}
is a frequently used data term in image processing, see \cite{Aujol2006,Liu2018}. 
The clean image $u$ is supposed to be piece-wise smooth. That is, it contains only few strong edges, which is reflected by overall small directional differences in all directions. In contrast, stripe artifacts are characterized by large values of the difference operators $\nabla_{x} s$ and $\nabla_{z} s$ orthogonal to the stripe direction and small values of the operator $\nabla_y s$ in direction parallel to it. This behaviour is visualized in Fig. \ref{fig:VarMethodsMotivation} and incorporated by the function
\begin{equation}
    \norm{\nabla_y s}_1 = \sum_{i,j,k} \vert (\nabla_y s)_{i,j,k}\vert.
    \label{eq:yStripes}
\end{equation}

For increased performance regularization is added through 
\begin{align}
    \norm{s}_1 &= \sum_{i,j,k} \vert s_{i,j,k}\vert,\label{eq:Stripes}\\
    \iota_{[0,1]^N}(u) &=\begin{cases}
                            0 & \text{if } u\in[0,1]^N,\\
                            \infty & \text{if } u\notin[0,1]^N.
                        \end{cases}\label{eq:Indicator}
\end{align}
$\ell_1$-regularization through \eqref{eq:Stripes} promotes sparsity in the stripe image. It reflects the generally small area affected by artifacts. 
The indicator \eqref{eq:Indicator} is included to guarantee the outcome to live in the same value range as the input. In particular, values remain non-negative. 
Hence, further post-processing such as adjusting contrast or the intensity profile becomes obsolete. The resulting optimization problem reads
\begin{equation}\label{eq:optimizationproblem}
    \underset{u+s=u_0}{\mathrm{argmin}}\ \mu_1\norm{\nabla u}_{2,1}+\norm{\nabla_y s}_1 + \mu_2\norm{s}_1+ \iota_{[0,1]^N}(u),
\end{equation}
with weighting parameters $\mu_1,\mu_2 >0$. Due to scaling invariances $\lambda\iota_{[0,1]^N}=\iota_{[0,1]^N}$ and $\mathrm{argmin}\, f(x) = \mathrm{argmin}\, \lambda f(x)$ for $\lambda >0$, only two weights $\mu_1,\mu_2$ are required. 

\subsubsection{Minimization} 
Convex optimization problems such as \eqref{eq:optimizationproblem} can be solved via the primal-dual gradient hybrid method with extrapolation of the dual variable (PDHGMp), see \cite{Burger2016}. 
Under mild assertions on the objective function, the sequence generated by PDHGMp converges to a solution, see \cite{Burger2016,Chambolle2011} for details.

The weighting parameters $\mu_1, \mu_2$ in \eqref{eq:optimizationproblem} provide adjustability to control different aspects of the stripe removal, see Supplement 1. The following rules of thumb can be established:
\begin{itemize}
    \item[(i)] $\mu_1$ controls the smoothness of the outcome. It strongly regulates the amount of stripe removal to be applied. Larger values will result in stronger removal of stripe artifacts. However, vertical and stripe-like image structures may become affected which results in a smoothed image.
    \item[(ii)] $\mu_2$ acts like a counterpart to $\mu_1$ and controls how strictly attention is paid to the stripe-like appearance of affected structures.  
    Increasing its value generally tends towards a lower reduction of artifacts and image structures. However, in combination with $\mu_1$ it enforces that for the most part only structures with concise stripe properties, e.g., thin, long and vertically aligned, are removed. 
    \item[(iii)] The ratio of $\mu_1$ and $\mu_2$  has a strong influence on the general outcome. Scaling both parameters by an equal factor changes the amount of stripe removal while keeping the effect on image structures mostly unchanged.
\end{itemize}

In addition to offering adjustability through two parameters, variational methods are generally less susceptible to stripes deviating from being thin and long which we will see in Section \ref{sc:Results}. Furthermore, they can easily be adapted to specific settings. For example, for 2D image data we replace the total variation \eqref{eq:TotalVariation} by its lower-dimensional counterpart 
\begin{equation}
    \norm{\nabla^{(2D)} u}_{2,1} = \sum_{i,j}\sqrt{(\nabla_{x} u)_{i,j}^2 + (\nabla_{y} u)_{i,j}^2}.
    \label{eq:TotalVariation2D}
\end{equation}
The lower spatial resolution of LSFM images along the displacement direction $z$ of the light-sheet  can be incorporated by using $\nabla^*_z u = \rho_z \nabla_z u$ with $0\leq\rho_z\leq 1$. This emphasizes that information along the $z$-direction is less reliable and coherent. The case $\rho_z=1$ corresponds to \eqref{eq:optimizationproblem} and $\rho_z=0$ to the 2D case. For oblique stripe artifacts as described in \cite{Liu2018} we replace $\nabla_y s$ with a suitable directional difference operator $\nabla_{\theta} s$ in \eqref{eq:yStripes}. Lastly, a simultaneous multi-directional stripe removal is possible using a sum of penalization terms $\norm{\nabla_{\theta_i} s}_1$ of different stripe directions $\theta_i$.

\begin{figure}[htbp]
    \centering
    \includegraphics[width=1\textwidth]{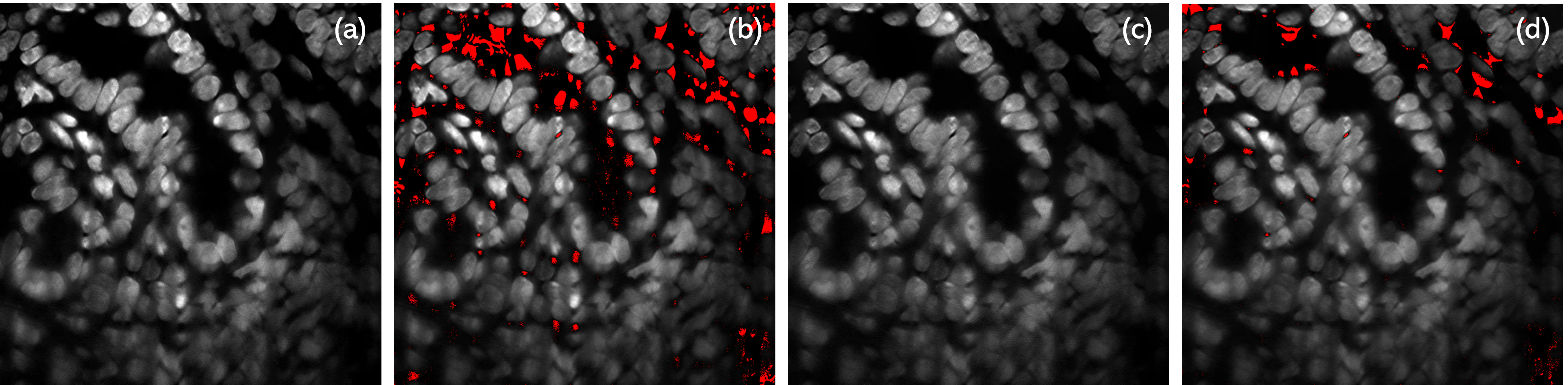}
    \caption{Stripe removal results using different variations of \eqref{eq:optimizationproblem} with $\mu_1=\frac{1}{3}$ and $\mu_2=\frac{1}{300}$. 2D with $\iota_{[0,1]^N}$ (a), 2D without $\iota_{[0,1]^N}$(b), 3D with $\iota_{[0,1]^N}$ (c) and 3D without $\iota_{[0,1]^N}$ (d). Areas with values less than 0 are colored in red and were clipped to 0 for comparability.}\label{fig:ProjvsNoProj}
\end{figure}

In Fig. \ref{fig:ProjvsNoProj} we visually compare the influence of processing in 2D and 3D and the difference of restricting optimization to the value range $[0,1]$ or not. 
At first glance, results are almost identical. However, at closer inspection some noticeable differences can be spotted. For example, when moving from 2D to 3D, we see a subtle increase in brightness in the background area left to the center. Inspection of the 3D image stack reveals that structural components can be found in the neighbouring slices which explains this phenomenon. Therefore, when interested in a single image slice it is sufficient to use the 2D variation. However, for 3D the incorporation of depth information ensures reasonably smooth intensity profiles and prevents sudden jumps. This becomes beneficial when further processing, such as image segmentation, is performed on the 3D image. 

The indicator function does not cause drastically different results if we clip the intensity profile to the same range after processing. However, as the affected areas are not negligible, we argue that removing the need for further steps is beneficial and prevents the introduction of errors.


\section{Results}\label{sc:Results}
In this section, we evaluate and compare the results of stripe removal methods. Therefore, we consider both, synthetic and real image data and evaluate performance on selected 2D slices.

\subsection{Performance Measures}
Performance of the destriping methods is evaluated by visual inspection and assisted by three quality metrics: the peak signal-to-noise ratio (PSNR), the multi-scale structural similarity index measure (MS-SSIM) \cite{Wang2003}, and the 'curtaining' metric proposed by Roldan \cite{Roldan2024}. PSNR is one of the most common quality metrics used in image processing. For any image $u\in[0,1]^{n_x\times\, n_y}$ and reference $u^*$
\begin{equation}
    \mathrm{PSNR}(u,u^*) = -10\log\underbrace{\left( \frac{1}{n_x n_y} \sum_{i=1}^{n_x}\sum_{j=1}^{n_y}\left[u(i,j) - u^*(i,j)\right]^2     \right)}_{=\text{ Mean squared error}}
\end{equation}
captures the power of corruptions in comparison to the reference as scalar value in $[0,\infty]$ where larger values correspond to higher quality. 

MS-SSIM is an extension of the popular structural similarity index measure 
\begin{equation}
    \mathrm{SSIM}(u,u^*) = l(u,u^*)\cdot c(u,u^*)\cdot \tilde{s}(u,u^*)
\end{equation}
based on measurements for luminance $l$, contrast $c$ and structure $\tilde{s}$. It attains values in $[0,1]$ with larger values corresponding to stronger similarity. To obtain the multi-scale version, measurements of different image scales are combined, see \cite{Wang2003}. 

Lastly, the curtaining metric from \cite{Roldan2024} measures the presence of stripes. It is based on the same ideas as Fourier filtering methods, see Fig. \ref{fig:FourierStripes}, and measures how strongly Fourier coefficients are condensed around a central horizontal line. The range of values is $[0,1]$ with larger values corresponding to better quality, i.e., less stripes.  Since this metric is based on similar ideas as Fourier filter methods, it is biased toward such and comparably sensitive to deviations from ideal stripes. However, in contrast to the prior measures, the curtaining metric does not require a reference image and is tailor-made for characterizing corruptions by stripes. 

\subsection{Description of employed stripe removal methods}
For our comparison we consider the following methods:
\begin{itemize}
    \item The Fourier based multi-directional stripe remover (MDSR) proposed by Liang et al. \cite{Liang2016} as described in Section \ref{ssc:FourierFiltering}. 
    The number of directions $n_{\mathrm{dir}}=8$ and depth of decomposition $n_{\mathrm{dec}}=5$ are fixed for all following results. Processing was performed using the 'Nonsubsampled Contourlet Toolbox' \cite{NSCTToolbox} in Matlab. In particular, the function 'nsctdec' with default options for generating filter banks was used. The standard deviation $\sigma$ of the Gaussian damping function and the damping fall-off parameter $\sigma_a$ remain as free parameters. 
    \item The variational stationary noise remover (VSNR) proposed by Fehrenbach et al. \cite{Fehrenbach2012}. 
    For our application, we chose elementary noise patterns $\psi_i$ as the real part of three differently sized Gabor filters, see Fig. \ref{fig:GaborPatterns}, such that short, medium and long stripes can be detected simultaneously. The corresponding objective function reads
    \begin{equation}
        \underset{\substack{u+s = u_0 \\ \norm{\lambda}_{\infty}\leq1}}{\mathrm{argmin}}\ \norm{\nabla u}_{1,\epsilon} + \alpha_1\norm{\lambda_1}_1 + \alpha_2\norm{\lambda_2}_1  + \alpha_3\norm{\lambda_3}_1\quad\text{s.t}\quad  s=\sum_{i=1}^3 \lambda_i\ast \psi_i,
    \end{equation}
    with free weighting parameters $\alpha_1,\alpha_2,\alpha_3$, $\epsilon >0$,
    \begin{align*}
        \norm{q}_{1,\epsilon} = \sum_{i,j} \phi_{\epsilon}\left(\sqrt{q_{i,j,1}^2 + q_{i,j,2}^2}\right), \quad 
        \phi_{\epsilon}(x)=\begin{cases}
            \frac{x^2}{2\epsilon} & \abs{x}\leq \epsilon, \\
            \abs{x} - \frac{\epsilon}{2} & \text{otherwise}.
        \end{cases}
    \end{align*}   
    We use the Matlab implementation \cite{weiss2024} provided by the author Pierre Weiss and performed 25000 steps of the PDHGMp optimization algorithm. 

    \begin{figure}[htbp]
        \centering
        \includegraphics[width=0.8\textwidth]{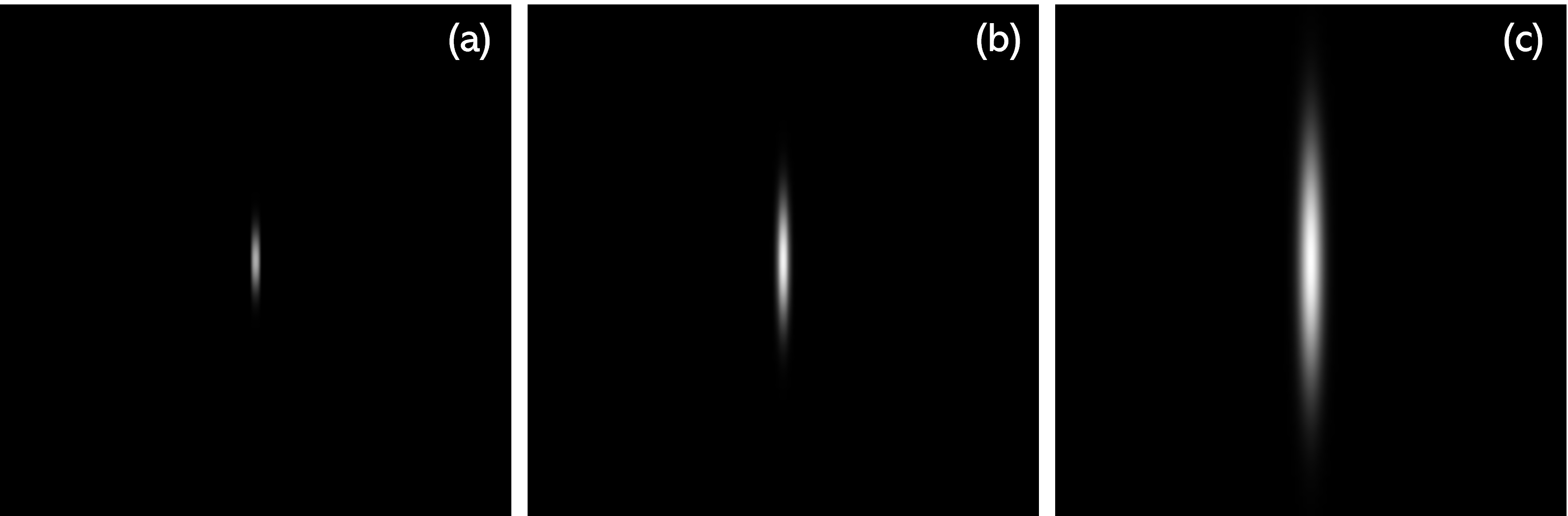}
        \caption{Elementary stripe patterns $\psi_i, i=1,2,3$ at 4x magnification.}\label{fig:GaborPatterns}
    \end{figure}

    \item The variational method \eqref{eq:optimizationproblem} in its 2D form which we will refer to as 'general stripe remover' (GSR). 
    Free parameters are $\mu_1$ and $\mu_2$. An approximation of the optimum is attained using 25000 iteration steps of the PDHGMp.
\end{itemize}

Our main objective for comparing methods was to achieve outcomes which are visually appealing and  optimal in terms of the metrics at the same time. However, the performance measures consider only some aspects of the image quality, e.g., the curtaining metric of a strongly smoothed image would suggest optimal image quality even if the actual image content was destroyed. Therefore, visual inspection is required to confirm numerical results. Initial parameters of the methods were obtained from a coarse grid-search and visual and numerical assessment. Afterwards, parameters were manually fine-tuned towards a reasonably optimal solution where any significant perturbation in parameters yielded either visual degradation, significant reductions of the quality metric values or both. 

For MDSR and VSNR the results were not clipped to values in $[0,1]$ when calculating the metrics to reflect the true outcome of the algorithms. However, for proper visualization clipping was applied. From each 3D image used in the following part, a representative slice with prominent stripe artifacts was selected. 

\subsection{Synthetic Data}\label{ssc:SyntheticData}
\begin{figure}[htbp]
    \centering
    \includegraphics[width=1\textwidth]{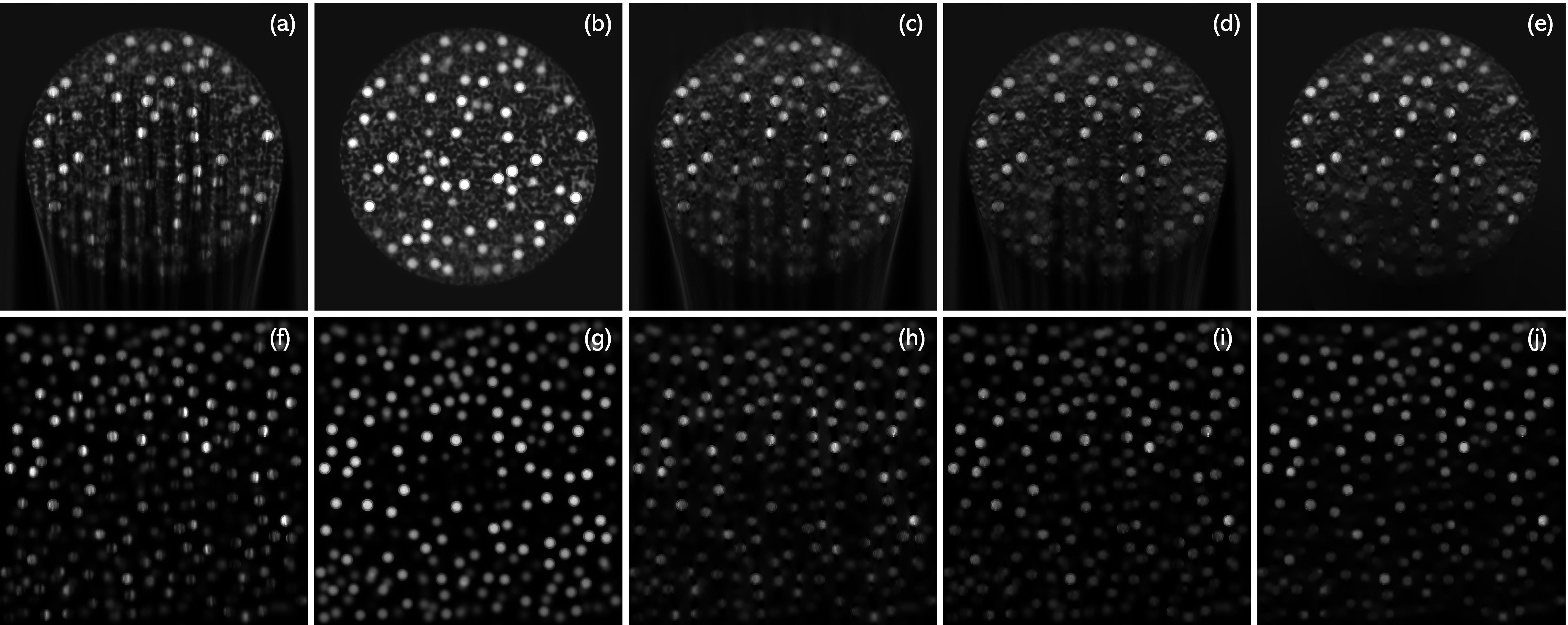}
    \caption{Synthetic image data and destriping results of a cell cluster (top) and dispersed cells (bottom). Input image (a), reference (b), MDSR, $\sigma=12$, $\sigma_a=0.3$ (c), VSNR, $\alpha=(3,5,10)$ (d), GSR, $\mu=(\frac{1}{3},\frac{1}{300})$ (e), input image (f), reference (g), MDSR, $\sigma=24$, $\sigma_a=0.3$ (h), VSNR, $\alpha=(3,5,10)$ (i), GSR, $\mu=(\frac{1}{2},\frac{1}{60})$ (j).}\label{fig:ArtificialResults}
\end{figure}

 Synthetic images are obtained by modeling the imaging setup in LSFM and simulating light transport \cite{Rohrbach2009} using the python package \textit{biobeam} \cite{Weigert2018}. The simulation allows for the retrieval of reference images without stripes by neglecting interactions of light and matter. However, this removes natural light attenuation which is usually visible as an intensity decrease along the illumination direction, see Fig. \ref{fig:ArtificialResults} (a-b) and (f-g). Hence, the reference images cannot be interpreted as ideal destriping results and a comparison should be done with caution. For details on the simulation and generation of synthetic images, see Supplement 1.  
The results shown in Fig. \ref{fig:ArtificialResults} display visually significant improvements in image quality across all stripe removal methods. GSR produces the strongest reduction of stripes including the trailing artifacts below the spherical body, see Fig. \ref{fig:ArtificialResults} (c). VSNR removes most thin stripe artifacts while wide stripes remain visible in Fig. \ref{fig:ArtificialResults} (d). Furthermore, faint smearing artifacts are introduced in cavities in Fig. \ref{fig:ArtificialResults} (i). MDSR yields the lowest reduction of stripes in both cases with more visible remainders and stronger smearing artifacts, see Fig. \ref{fig:ArtificialResults} (c) and (h). PSNR and MS-SSIM (Table \ref{tab:ArtificialResults}) do not change significantly for all methods. The curtaining metric increases significantly for all methods on the cell cluster image but only for GSR on the dispersed cells. 

\begin{table}[htbp]
    \centering
    \begin{tabular}{ccccccc}\hline
        & \textbf{Metric} & \textbf{Input} & \textbf{Reference} & \textbf{MDSR} & \textbf{VSNR} & \textbf{GSR} \\\hline
        \multirow{3}{*}{Cell Cluster} & Curtaining & 0.78 & 0.90 & \cellcolor{green!25}0.99 & 0.90 & 0.97 \\
        &PSNR & 24.80 & $\infty$ & \cellcolor{green!25}24.64 & 24.00 & 24.58 \\
        &MS-SSIM & 0.86 & 1 & 0.85 & 0.84 & \cellcolor{green!25}0.87 \\\hline
        \multirow{3}{*}{Dispersed Cells} & Curtaining & 0.91 & 0.96 & 0.91 & 0.91 & \cellcolor{green!25}1 \\
        &PSNR & 28.22 & $\infty$ & \cellcolor{green!25}27.25 & 27.13 & 27.20 \\
        &MS-SSIM & 0.88 & 1 & 0.84 & 0.84 & 0.84 \\\hline
    \end{tabular}
    \caption{Quality metrics of the images in Fig. \ref{fig:ArtificialResults}. Larger is better. The best result is highlighted in green.}\label{tab:ArtificialResults}
\end{table}

\subsection{LSFM images of biological samples}\label{ssc:RealData}
\begin{figure}[htbp]
    \centering
    \includegraphics[width=1\textwidth]{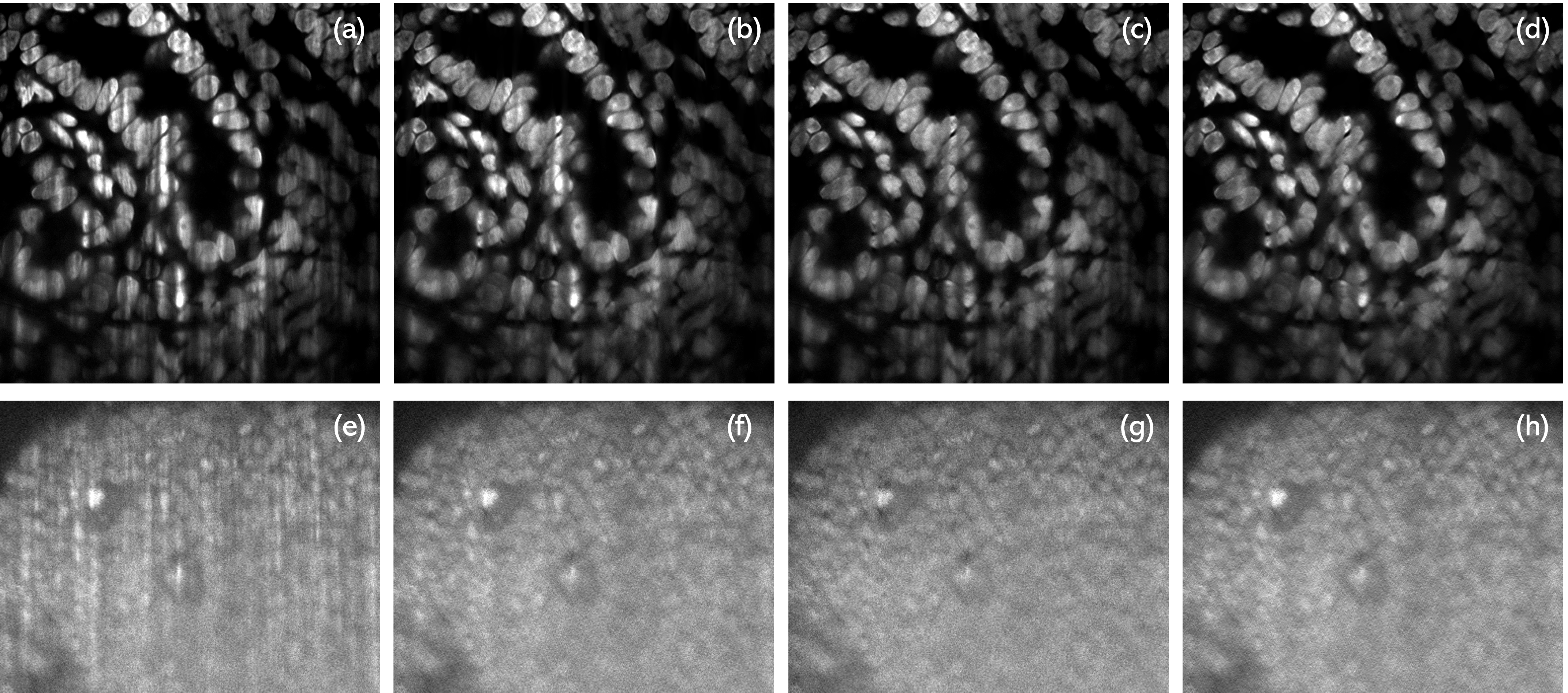}
    \caption{Results on LSFM images. Mouse intestine cells (top) and HeLa-cluster (bottom). In put image (a), MDSR, $\sigma = 12$, $\sigma_a=0.3$ (b), VSNR, $\alpha=(3,5,10)$ (c), GSR, $\mu=(\frac{1}{3},\frac{1}{300})$ (d), input image (e), MDSR, $\sigma = 5$, $\sigma_a=0.3$ (f), VSNR, $\alpha=(10,5,1)$ (g), GSR, $\mu=(\frac{1}{3},\frac{1}{300})$ (h). Corresponding curtaining values are 0.58 (a), 0.84 (b), 0.78 (c), 0.95 (d),  0.24 (e), 0.90 (f), 0.99 (g) and 0.96 (h).}\label{fig:LSFMResults}
\end{figure}

We consider two LSFM images with different appearances. Fig. \ref{fig:LSFMResults} (a-d) shows an organoid of cultivated mouse intestine cells where individual cells are clearly visible and separated. Fig. \ref{fig:LSFMResults} (e-h) displays a cluster of HeLa cells which has a more homogeneous appearance.

As for the synthetic data, GSR yields the strongest reduction in stripes including wide and slightly oblique artifacts. However, for the mouse intestine cells some faint smearing artifacts were introduced in structural cavities, see Fig. \ref{fig:LSFMResults} (d), and few wide stripes remain visible in Fig. \ref{fig:LSFMResults} (h). VSNR removes all visible stripes in the HeLa-cluster but is unable to remove wide stripes for the mouse intestine cells, see Fig. \ref{fig:LSFMResults} (g) and (c). In the latter, thin but oblique artifacts are reduced but remain visible and very faint smearing artifacts were introduced similar to GSR. The results by MDSR are comparable to VSNR on the mouse intestine cells. Some smearing artifacts are introduced but there are even more remainders of thin oblique stripes than for VSNR, compare Fig. \ref{fig:LSFMResults} (b-c). In the HeLa-cluster barely any artifacts remain, similar to the result of GSR. 

\subsection{Application for other imaging methods}
\begin{figure}[htbp]
    \centering
    \includegraphics[width=\textwidth]{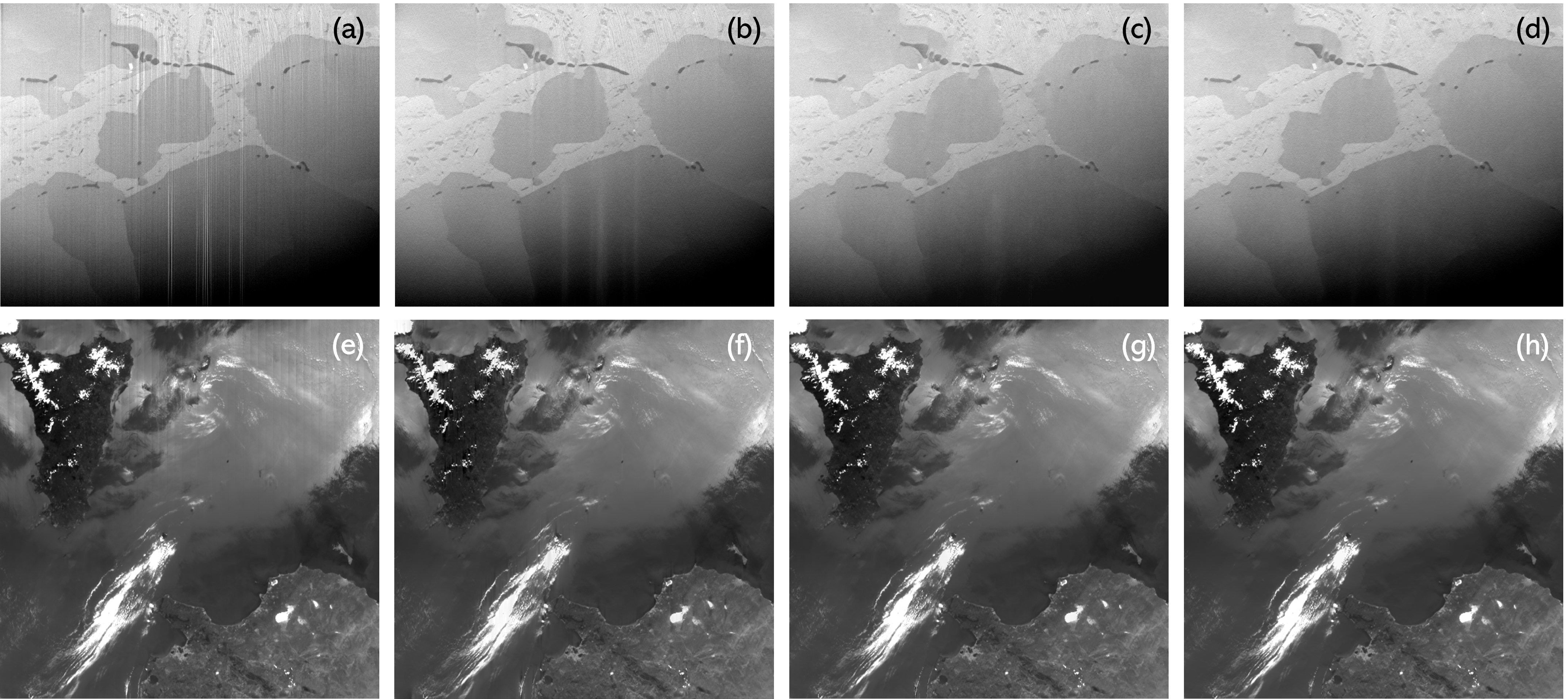}
    \caption{Results on FIB-SEM (top) and Terra MODIS (bottom) examples. Input image (a), MDSR, $\sigma = 12$, $\sigma_a=0.3$ (b), VSNR, $\alpha=(2,10,2)$ (c), GSR, $\mu=(\frac{2}{5},\frac{1}{300})$ (d), input image (e), MDSR, $\sigma = 10$, $\sigma_a=0.3$ (f), VSNR, $\alpha=(3,5,5)$ (g), GSR, $\mu=(\frac{7}{30},\frac{1}{300})$ (h). Corresponding curtaining values are 0.18 (a), 0.54 (b), 0.95 (c), 1 (d), 1 (e), 1 (f), 1 (g) and 1 (h).}\label{fig:NonLSFMResults}
\end{figure}

We also consider data by other imaging methods where stripe artifacts are commonly encountered. Our examples shown in Fig. \ref{fig:NonLSFMResults} include a FIB-SEM image of a tin bronze cast alloy and a Terra MODIS satellite image. Both display more uniform thin and long stripes than the LSFM images.  

GSR removes stripes entirely from the Terra MODIS image and leaves only few faint remainders of stripes in the FIB-SEM example. Similar results can be observed for VSNR but remainders in the FIB-SEM image are slightly stronger. MDSR introduces visible artifacts in high contrast areas in the Terra MODIS image despite removing all stripes. Additionally, it leaves visible remainders of wide stripes in the FIB-SEM image where multiple stripes occurred close together. 
    
\subsection{Discussion}
The results shown in Section \ref{sc:Results} reveal that GSR outperforms existing methods such as MDSR and VSNR. It produces consistently good results in terms of stripe removal quality and retention of image structures while offering adjustability through the weighting parameters $\mu_1$ and $\mu_2$. 

The synthetic data which offer stripe-free reference images showed that only GSR is able to remove stripes of larger widths and slightly oblique artifacts such as the trails in Fig. \ref{fig:ArtificialResults} (a-e). This is explained by the fact that the directional differences in functional \eqref{eq:optimizationproblem} also penalize wide and oblique stripe artifacts. 
On the contrary, VSNR relies heavily on pre-selected patterns such that the removal of oblique artifacts requires oblique patterns to work well. Similarly, MDSR assumes that stripe information condenses closely into a horizontal coefficient band in the Fourier domain. Deviations in stripe direction will yield a significantly lower reduction of stripe information.

The quality metrics in Table \ref{tab:ArtificialResults} confirm visual observations for the most part. The high curtaining values produced by MDSR can be attributed to bias as both rely on the same theory and properties of Fourier space. Furthermore, the lack of improvement of MDSR and VSNR on the dispersed cells image is caused by the introduction of smearing artifacts. Note that PSNR and MS-SSIM do not improve across all methods. There are multiple reasons that explain this phenomenon. First, the removal of stripe artifacts results in a noticeable decrease in contrast, specifically in areas which were strongly corrupted. Furthermore, the reference image has an overall higher brightness and more details because light attenuation and interactions with matter are neglected during generation. Hence, a full restoration of the original structures is neither anticipated nor possible. Despite of these problems, the fact that PSNR and MS-SSIM have not decreased suggests that structural improvements by stripe reduction have compensated any side effect of the removal process.  

For the real image data we observe similar results with GSR producing consistently good results. Specifically, the amount of stripe reduction on the mouse intestine cells shown in Fig. \ref{fig:LSFMResults} is impressive as almost all artifacts are removed. This showcases the strengths of GSR as the image is corrupted by a broad range of different stripe artifacts, including various widths, lengths, intensities and deviations in orientations. However, we observe that there are some problems with introducing smearing artifacts in stripe-like cavities of the structure. This can be avoided by adjusting $\mu_1$ and $\mu_2$ at the cost of lower stripe reduction.   

VSNR develops its full potential on the HeLa-cluster. The result is close to perfect showing no visible artifacts while retaining good contrast and image details. This is achieved since the image reflects an ideal scenario where stripe patterns perfectly fit assumptions on the stripe artifacts, structures have low contrast and a uniformly colored background. The latter hides smoothing artifacts that may be introduced in the process. The results by GSR on this image are on par with slightly better contrast and detail preservation but few large and smooth stripes remain visible. This is most likely caused by a lower smoothing in comparison to VSNR.  

The performance of MDSR is somewhat disappointing. MDSR yields the overall lowest stripe reduction while introducing the most severe artifacts across all examples. In particular, the results on the Terra MODIS example is surprising as the thin and periodic stripes should be the ideal setting for this method. However, the high contrast in certain areas seems to produce unforeseen problems. Despite all criticism, the result on the HeLa-cluster shows that performance similar to VSNR and GSR is achievable under the right circumstances. In general, MDSR is more sensitive to stripes deviating from being thin, long and vertically oriented than its variational competitors which is caused by the corresponding influence on stripe information in the Fourier domain.

\section{Conclusion}
In this paper, we evaluated several stripe removal methods on LSFM, FIB-SEM and remote sensing images. We explained Fourier filtering and variational methods using selected examples. Extensions of existing propositions increased the quality of stripe removal. We used synthetically generated LSFM data and real images to evaluate and compare the performance of stripe removal by visual inspection. The observations are confirmed by quality metrics.  
Our proposed objective function \eqref{eq:optimizationproblem}, later referred to as GSR, produced consistent results with better stripe removal than prior propositions.  
This includes stripes of varying lengths, widths, intensities and slightly deviated directions. Other methods such as the previously published MDSR and VSNR produce a lower reduction of artifacts and show inconsistencies when facing different image appearances and structures.

The usage of synthetic LSFM data which offer a stripe-free reference image is a novel supplement of real data for comparing quality of stripe removal. It provides valuable insights into the capabilities and limitations of stripe removal method, e.g., inconsistent shape reconstruction, insufficient stripe removal or the introduction of smearing artifacts. Furthermore, it allows for the use of standard image quality metrics such as PSNR and MS-SSIM which further assist in a precise assessment. The recently proposed curtaining metric proved to be an valuable indicator of stripe corruption. Its major advantage is that it is calculated directly on the image and does not require a reference. However, it does not replace the need for visual inspection.

\section*{Supplemental document} See Supplement 1 for supporting content.

\section*{Funding}
Federal Ministry of Education and Research (BMBF), Project: Synthetic Data for Machine Learning Segmentation of Highly Porous Structures from FIB-SEM Nano-tomographic Data (poSt), Funding number: 01IS21054A

\section*{Acknowledgments} We thank Martin Weigert for guidance on installing biobeam and its application for modelling LSFM imaging as well as Michael Engstler for providing the FIB-SEM image data. Furthermore, we want to thank Diego Rold\'an for providing and explaining his curtaining metric and Jannik Reiser for his help with data processing. 

\section*{Disclosures}
The authors declare no conflicts of interest.

\section*{Data Availability Statement}
The methods presented in this paper will be made available. This includes MATLAB code for all used methods and scripts for processing images. Furthermore, a Jupyter-notebook for generating synthetic LSFM data and a txt-file for creating the Python environment on Windows 11 are provided.

\bibliography{Sources-Finale}

\end{document}